\documentclass[reprint,aps,prb,amsmath,amssymb]{revtex4-1}

\usepackage[]{graphicx}
\usepackage[]{units}
\usepackage{times}
\usepackage{bm}
\usepackage{ulem}		
\usepackage{color}

\newcommand{\comment}[1]{}

\renewcommand{\emph}{\textit}

\begin{document}

\title{Theory of optically induced F\"orster Coupling in van der Waals coupled Heterostuctures}
\author{Malte Selig$^{1,2}$}
\author{Ermin Malic$^2$}
\author{Kwang Jun Ahn$^3$}
\author{Norbert Koch$^4$}
\author{Andreas Knorr$^1$}
\affiliation{$^1$Nichtlineare Optik und Quantenelektronik, Institut f\"ur Theoretische Physik, Technische Universit\"at Berlin,  10623 Berlin, Germany}
\affiliation{$^2$Chalmers University of Technology, Department of Physics, SE-412 96 Gothenburg, Sweden}
\affiliation{$^3$Department of Energy Systems Research, Ajou University, 16499 Suwon, South Korea}
\affiliation{$^4$ Institut f\"ur Physik and IRIS Adlershof, Humboldt-Universit\"at zu Berlin, 12489 Berlin, Germany}

\begin{abstract}

We investigate the impact of optically induced F\"orster coupling in van der Waals heterostructures consisting of graphene and a monolayer transition metal dichalcogenide (TMD). In particular, we predict the corresponding dephasing rates and a fast energy transfer between the TMD layer and graphene being in the picosecond range. Exemplary we find a transition rate of thermalized excitons of about \unit[4]{ps$^{-1}$ in a MoSe$_2$-graphene stack at room temperature. This timescale is in good agreement with the recently measured exciton lifetime in this heterostructure.}
\end{abstract}

\maketitle

\begin{figure}[t!]
 \begin{center}
\includegraphics[width=0.9\linewidth]{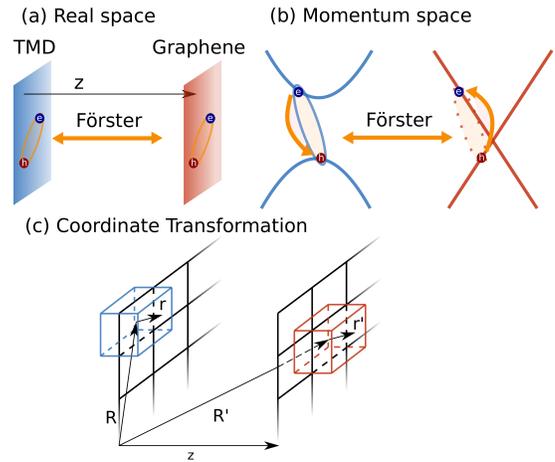}
 \end{center}
 \caption{\textbf{Schematic illustration of the F\"orster coupling.} (a) Real space view: a TMD and a graphene layer with a distance $z$ to each other. F\"orster coupling between the layers leads to an energy transfer. (b) Momentum space view: excitons in the TMD and in the graphene couple via the F\"orster mechanism under conservation of energy and momentum. (c) Illustration of the coordinate transformation.}
 \label{Schema}
\end{figure}

Since the discovery of graphene in 2004, a new research field on atomically thin quasi two dimensional (2D) materials has been established. In particular, monolayers of transition metal dichalcogenides (TMDs) became one of the most investigated materials beyond graphene. They exhibit technologically promising characteristics, such as a direct band gap, tightly bound electron-hole pairs (excitons)\cite{Mak2010,Ramasub2012,Chernikov2014}, strong light-matter interaction\cite{Li2014,Moody2015},  and valley-selective circular dichroism \cite{Cao2012,Schmidt2016}. 
Recent progress in growth techniques has also enabled the production of van der Waals heterostructures by vertically stacking atomically thin monolayers \cite{Shi2012,Geim2013}. Consequently, heterostructures consisting of graphene and a monolayer TMD have been investigated theoretically \cite{Trushin2017} and experimentally \cite{Georgiou2012,Britnell2013,HeJiaqi2014,Lin2014,WZhang2014,Hill2017,Froehlicher2018}. Performing differential reflection measurements, J. He and co-workers found the relaxation of optically injected carriers from tungsten disulfide to graphene to occur within \unit[1]{ps}\cite{HeJiaqi2014}. In recent photoluminescence measurements G. Froelicher and co-workers found found a shortened exciton lifetime of about \unit[1]{ps} in a coupled MoSe$_2$-graphene structure at room temperature \cite{Froehlicher2018}. In another study, Hill and co-workers found an additional broadening of \unit[5]{meV} and a redshift of \unit[23]{meV} of the excitonic resonance in a linear optics experiment due to the coupling between monolayer WS$_2$ and graphene\cite{Hill2017}. The experimental data have not been complemented by microscopic theory yet: Possible coupling mechanisms constitute of electronic tunneling, Dexter- and F\"orster energy transfer \cite{Richter2006,Rozbicki2008,Specht2015,Ovesen2018}.
Since both materials exhibit strong optical dipole moments, F\"orster coupling is expected to have a significant impact on the excitation transfer in these heterostructures, if the barrier potential is large enough to suppress electronic overlap \cite{Specht2015}, cf. Fig. \ref{Schema}. 

F\"orster-induced relaxation dynamics were investigated recently in heterostructures of different dimensions, namely between quantum dots\cite{Richter2006,Rozbicki2008}, between graphene and attached molecules\cite{Malic2014} and between two quantum wells\cite{Batsch1993}.
Here, we present a microscopic approach based on the Heisenberg equation of motion formalism allowing us to derive an analytic description of the F\"orster coupling in a van der Waals heterostructure consisting of graphene and a TMD monolayer. We predict the impact of the F\"orster coupling on the optical response of the heterostructure including the linewidth and spectral shift of excitonic resonances. 
In principle, also charge transfer (tunneling and Dexter processes) between both monolayers could affect the spectral width and the spectral position of the exciton line in the TMD monolayer. However, charge transfer does strongly dependent on the overlap of the wavefunctions of electrons in the TMD and graphene. For electronically decoupled monolayers, the F\"orster coupling can be expected to dominate tunnel and Dexter coupling \cite{Specht2015}. For closely stacked heterostuctures however further investigation are required to evaluate the actual strength of the Dexter- and tunnel- coupling. This requires first principle methods and is addressed in future work. On the other hand, pure F\"orster coupling provides at least a limiting case that can be discussed in the analysis of experiments without or involving electronic overlap.

\section{Theoretical Model}

We start with defining the many-particle Hamilton operator describing the Coulomb interaction between an exemplary TMD layer and graphene
\begin{equation}
H= \sum_{\mathbf{k},\mathbf{k'},\lambda,\lambda'}\sum_{\mathbf{q},\mathbf{q'},\nu,\nu'} V^{\lambda \nu \lambda' \nu'}_{\mathbf{k}\mathbf{q} \mathbf{k'} \mathbf{q'}} a^{\dagger \lambda}_\mathbf{k} a^{\dagger \nu}_\mathbf{q} a^{\nu'}_\mathbf{q'} a^{\lambda'}_\mathbf{k'},
\end{equation}
with electron annihilation (creation) operators $a^{(\dagger) \lambda}_\mathbf{k}$ acting on states with the band index $\lambda=c,v$ and the momentum $\mathbf{k}$. We denote electrons in the TMD monolayer by $(\mathbf{k}^{(')},\lambda^{(')})$ and in graphene by $(\mathbf{q}^{(')},\nu^{(')})$. The appearing Coulomb matrix element inducing the F\"orster transfer reads 
\begin{equation}
V^{\lambda \nu \lambda' \nu'}_{\mathbf{k}\mathbf{q} \mathbf{k'} \mathbf{q'}}=\langle \Psi^{*\lambda}_\mathbf{k}(\mathbf{r}) \Psi^{*\nu}_\mathbf{q}(\mathbf{r'}) V(\mathbf{r}-\mathbf{r'}) \Psi^{\nu'}_\mathbf{q'}(\mathbf{r'}) \Psi^{\lambda'}_\mathbf{k'}(\mathbf{r})\rangle
\end{equation}
with the single particle electronic wavefunctions $\Psi_\mathbf{k}^{\lambda}$. We consider both materials to be aligned in the x-y-plane. Further, $V(\mathbf{r}-\mathbf{r'})=\frac{e^2}{4 \pi \epsilon_0 \epsilon} \frac{1}{|\mathbf{r}-\mathbf{r'}|}$ denotes the three dimensional Coulomb potential, with $\epsilon_0$ the vacuum permittivity and $\epsilon$ the mean dielectric constant of the surrounding material. This assumption is valid, since it differs only weakly from the exact static Coulomb potential for a dielectric surrounding typically found in experiments \cite{Hill2017}, i.e. a TMD monlayer on a quartz substrate. This was checked carefully by explicitly evaluating the Possion equation \cite{Ovesen2018}.
Next, we decompose the space coordinates $\mathbf{r}=\mathbf{R_n}+\mathbf{r_n}$ in both constituents in $\mathbf{R_n}$ pointing to the center of the n-th unit cell (uc) and $\mathbf{r_n}$ describing a point  within the unit cell. Thus, the integral transforms to $\int_{\mathbb{R}^3} d^3 r =\sum_\mathbf{R_n} \int_{\text{uc}} d^3 \mathbf{r_n}$. Here, the sum over the unit cells $\mathbf{R_n}$ is restricted to the monolayers, whereas the $\mathbf{r_n}$ integrations are performed in the 3 dimensional unit cells, cf. figure \ref{Schema} (c).
Now, we perform a Taylor expansion for the intra cell coordinate $\mathbf{r_n}$ and assume $|\mathbf{r_n-r_n'}| \ll |\mathbf{R_n-R_n'}|$ which can only be fulfilled, if the wavefunctions of the electrons in graphene and the TMD layer do not overlap. Here, $min|\mathbf{R_n}-\mathbf{R_n'}|=z$ denotes the distance between both constituents.  The monopole-monopole contribution of the Taylor expansion can be shown to contribute to the diagonal part of the Hamiltonian, providing an energy renormalization with respect to the uncoupled TMD-graphene heterostructure \cite{Richter2006}. This energy renormalization vanishes in the linear optics limit and is therefore neglected in the following. The monopole-dipole can be neglected within the rotating-wave approximation of excitation with optical frequencies \cite{Specht2015} and therefore the dominating part is the dipole-dipole contribution of the Coulomb potential
$
\frac{1}{|\mathbf{R_n}+\mathbf{r_n}-\mathbf{R_n'}-\mathbf{r_n'}|}=
\frac{\mathbf{r_n}\cdot\mathbf{r_n'}}{|\mathbf{R_n}-\mathbf{R_n'}|^3}
-3\frac{\mathbf{r_n}\cdot(\mathbf{R_n}-\mathbf{R_n'})\mathbf{r_n'}\cdot(\mathbf{R_n}-\mathbf{R_n'})}{|\mathbf{R_n}-\mathbf{R_n'}|^5}
.$

Assuming electronic Bloch functions in both structures, we obtain for the coupling element 
\begin{align}
&V^{\lambda \nu \lambda' \nu'}_{\mathbf{k}\mathbf{q} \mathbf{k'} \mathbf{q'}}=\frac{\Omega^2}{4\pi \epsilon_0 \epsilon A^2} \sum_\mathbf{R_n,R_n'} e^{-i \mathbf{R_n} (\mathbf{k}-\mathbf{k'})} e^{-i \mathbf{R_n'} (\mathbf{q}-\mathbf{q'})} \times\nonumber  \\
&\times \left(\frac{\mathbf{d}^{\lambda \lambda'}_\mathbf{k k'}\cdot \mathbf{d}^{\nu \nu'}_\mathbf{q q'}}{|\mathbf{R_n}-\mathbf{R_n}'|^3}-3\frac{\mathbf{d}^{\lambda \lambda'}_\mathbf{k k'}\cdot(\mathbf{R_n}-\mathbf{R_n'})\mathbf{d}^{\nu \nu'}_\mathbf{q q'}\cdot(\mathbf{R_n}-\mathbf{R_n'})}{|\mathbf{R_n}-\mathbf{R_n'}|^5}\right).
\end{align}

$\mathbf{d}^{\lambda \lambda'}_\mathbf{k k'}= \frac{e}{\Omega}\int_{\text{uc}} d^3 r\, u^{* \lambda}_\mathbf{k}(\mathbf{r}) \,\mathbf{r}\, u^{\lambda'}_\mathbf{k'} (\mathbf{r})$ denotes the dipole moment with the lattice-periodic Bloch factor $u^{\lambda}_\mathbf{k}(\mathbf{r})$ and the area of the unit cell $\Omega$. The Bloch factors further contain envelope functions ensuring the confinement in $z$-direction. Since they are normalized with respect to the $z$-integration, they drop in the computation.

To further evaluate  the sums over  $\mathbf{R_n}^{(')}$, we introduce center-of-mass coordinates $\mathbf{S}=\frac{1}{2}(\mathbf{R_n}+\mathbf{R_n'}+\mathbf{z})$ and $\mathbf{s}=\mathbf{R_n}-\mathbf{R_n'}-\mathbf{z}$, with $\mathbf{z}$ being the vector pointing from the TMD layer to the graphene layer.
Then, we obtain the matrix element:
\begin{align}
&V^{\lambda \nu \lambda' \nu'}_{\mathbf{k}\mathbf{q} \mathbf{k'} \mathbf{q'}}=\frac{ \Omega}{4\pi \epsilon_0 \epsilon A} \sum_\mathbf{s,Q} e^{-i \mathbf{Q}\cdot(\mathbf{s}+\mathbf{z})}\delta_\mathbf{Q,k-k'}\delta_\mathbf{Q,q-q'}\times  \\ \nonumber
&\times \bigg(\frac{\mathbf{d}^{\lambda \lambda'}_\mathbf{k \, k-Q}\cdot \mathbf{d}^{\nu \nu'}_\mathbf{q \, q+Q}}{|\mathbf{s}+\mathbf{z}|^3}-3\frac{\mathbf{d}^{\lambda \lambda'}_\mathbf{k \, k-Q}\cdot(\mathbf{s}+\mathbf{z})\mathbf{d}^{\nu \nu'}_\mathbf{q \, q+Q}\cdot(\mathbf{s}+\mathbf{z})}{|\mathbf{s}+\mathbf{z}|^5}\bigg).
\end{align}
Without loss of generality, we write the dipole moments as $\mathbf{d}^{\lambda \lambda'}_\mathbf{k,k-Q} = d^{\lambda \lambda'}_\mathbf{k,k-Q} \mathbf{e}^{\lambda \lambda'}_1$ and $\mathbf{d}^{\nu \nu'}_\mathbf{q,q+Q} = d^{\nu \nu'}_\mathbf{q,q+Q} \mathbf{e}^{\nu \nu'}_2$ with $\mathbf{e}_{1/2}$ denoting the directions of the dipole moments in the TMD layer and in graphene, respectively. Before we provide the final Hamiltonian for the F\"orster coupling, we evaluate the sums over the electronic bands. We find, that contributions with $\lambda = \lambda'$ and $\nu = \nu'$ do not induce an interlayer energy transfer but an energy renormalization \cite{Specht2015}. This renormalization vanishes in the limit of linear optics and therefore is dropped from the Hamiltonian. Additionally we perform a rotating frame approximation to remove non-energy conserving terms for excitations with optical frequencies from the Hamiltonian. Thus we obtain the Hamiltonian
\begin{align}
&H_{F}=\sum_{\mathbf{k,q,Q}} \frac{1}{4\pi \epsilon_0 \epsilon A} d^{c v }_{\mathbf{k}+\frac{1}{2}\mathbf{Q}\,k-\frac{1}{2}\mathbf{Q}} d^{v c}_{\mathbf{q}-\frac{1}{2}\mathbf{Q}\,q+\frac{1}{2}\mathbf{Q}} a(\mathbf{Q},\mathbf{z}) \times \nonumber \\
&\times a^{\dagger c}_{\mathbf{k}+\frac{1}{2}\mathbf{Q}} a^{\dagger v}_{\mathbf{q}-\frac{1}{2}\mathbf{Q}} a^{ c}_{\mathbf{q}+\frac{1}{2}\mathbf{Q}} a^{ v}_{\mathbf{k}-\frac{1}{2}\mathbf{Q}} \, +h.c.\,
\end{align}
with $
a(\mathbf{Q},\mathbf{z})=\int d^2 s \frac{e^{-i \mathbf{Q}\cdot (\mathbf{s}+\mathbf{z})}}{|\mathbf{s}+\mathbf{z}|^5} (|\mathbf{s}+\mathbf{z}|^2 \mathbf{e}^{cv}_1 \cdot \mathbf{e}^{vc}_2-3 \mathbf{e}^{cv}_1 \cdot (\mathbf{s}+\mathbf{z}) \mathbf{e}^{vc}_2 \cdot (\mathbf{s}+\mathbf{z}) ).
$

The next step is to introduce electron-hole pair polarizations $P^{\dagger cv}_{\mathbf{k}+\frac{1}{2}\mathbf{Q},\mathbf{k}-\frac{1}{2}\mathbf{Q}}=a^{\dagger c}_{\mathbf{k}+\frac{1}{2}\mathbf{Q}} a^{v}_{\mathbf{k}-\frac{1}{2}\mathbf{Q}}$ in the TMD and $R^{\dagger cv}_{\mathbf{q}+\frac{1}{2}\mathbf{Q},\mathbf{q}-\frac{1}{2}\mathbf{Q}}=a^{\dagger c}_{\mathbf{q}+\frac{1}{2}\mathbf{Q}} a^{v}_{\mathbf{q}-\frac{1}{2}\mathbf{Q}}$ in graphene to further simplify the Hamilton operator. 
Performing a transformation to center-of-mass coordinates and projecting the relative coordinate to exciton wavefunctions $\varphi^{\kappa}_\mathbf{q}$ in graphene and $\varphi^{\mu}_\mathbf{k}$ in the TMD layer with quantum numbers $\kappa$ and $\mu$, we obtain for the Hamiltonian
\begin{align}
H_{F}=&\sum_{\lambda,\mu,\kappa,\mathbf{k,q,Q}} \frac{1}{4\pi \epsilon_0 \epsilon A} d^{cv}_{\mathbf{k}+\frac{1}{2}\mathbf{Q}\,k-\frac{1}{2}\mathbf{Q}} d^{vc}_{\mathbf{q}-\frac{1}{2}\mathbf{Q}\,q+\frac{1}{2}\mathbf{Q}} a(\mathbf{Q},\mathbf{z}) \times \nonumber \\
&\times \varphi^{* \mu}_\mathbf{k} \varphi^{\kappa}_\mathbf{q} P^{\dagger \mu}_\mathbf{Q} R^{\kappa}_\mathbf{Q}\, +h.c.\,.
\end{align}
Now, we assume that the dipole moments do not depend on the momentum at the band minimum ($d^{cv}_{\mathbf{k}+\frac{1}{2}\mathbf{Q}\,k-\frac{1}{2}\mathbf{Q}}\equiv d_T$ and $d^{vc}_{\mathbf{q}-\frac{1}{2}\mathbf{Q}\,q+\frac{1}{2}\mathbf{Q}} \equiv d_G$) and exploit the fact that at optical frequencies there are no bound excitons in graphene, thus we approximate  $\varphi^{\kappa}_\mathbf{q}=\delta^{\kappa}_\mathbf{q}$ in the free particle limit. In contrast, for TMDs with strongly bound excitons, we use the relation $\sum_\mathbf{k} \varphi^{\mu}_\mathbf{k}= \sqrt{A} \varphi^{\mu}(\mathbf{r}=0)$ (where $\varphi^{\mu}(\mathbf{r})$ is the TMD exciton wavefunction in realspace). The TMD wavefunction is obtained as eigenvector of the Wannier equation \cite{Kochbuch,Berghauser2014}, where we treated the appearing Coulomb potential within the Keldysh approach to account for the finite width of the TMD monolayer\cite{Keldysh1978,Berkelbach2013}. We obtain the F\"orster coupling Hamilton operator
\begin{align}
H_{F}=\sum_{\mu,\mathbf{q},\mathbf{Q}} V(\mathbf{Q},\mathbf{z},\mu) P^{\dagger \mu}_\mathbf{Q} R^{\mathbf{q}}_\mathbf{Q}\,+\,h.c. \label{Ham_Foe}
\end{align}
with  the center of mass dependent F\"orster coupling element
 $V(\mathbf{Q},\mathbf{z},\mu)= \frac{ a(\mathbf{Q},\mathbf{z}) d_T d_G \varphi^{\mu}(\mathbf{r}=0)}{4 \pi \epsilon_0 \epsilon \sqrt{A}}$. The F\"orster Hamiltonian, eq. \ref{Ham_Foe}, can be interpreted as the annihilation of a bound exciton in the TMD monolayer and the creation of a free electron hole pair in the graphene monolayer, cf. figure \ref{Schema} (b).

Now, we have all ingredients to define the Bloch equations for the exciton polarizations $P^{\nu}_\mathbf{Q}, R^{\mathbf{q}}_\mathbf{Q}$  by exploiting the Heisenberg equation of motion
\begin{align}
i \hbar \partial_t P^{\nu}_\mathbf{Q} &=E_\mathbf{Q}^{\nu} P_\mathbf{Q}^{\nu} + \sum_\mathbf{q} V(\mathbf{Q},\mathbf{z},\nu) R^{\mathbf{q}}_\mathbf{Q}, \label{GrapheneCoherence} \\
i \hbar \partial_t R^{\mathbf{q}}_\mathbf{Q} &=E_\mathbf{Q}^{\mathbf{q}} R_\mathbf{Q}^{\mathbf{q}} + \sum_\mu V(\mathbf{Q},\mathbf{z},\mu) P^{\mu}_\mathbf{Q}.\label{ExcitonicCoherence}
\end{align}
The first term in both equations accounts for the free energy of the polarizations  $E^{\nu}_\mathbf{Q}=E^{\nu} + \frac{\hbar^2 \mathbf{Q}^2}{2 M}$ with $E^{\nu}$ being the spectral position of the TMD exciton and $M=m_e+m_h$ as the total mass of the exciton. In graphene, we assume the energy to be independent of the center-of-mass momentum $E_\mathbf{Q}^{\mathbf{q}}=\hbar v_F (|\mathbf{q}+\frac{1}{2}\mathbf{Q}|+|\mathbf{q}-\frac{1}{2}\mathbf{Q}|)\approx 2 \hbar v_F |\mathbf{q}|$ with $v_F$ as the Fermi velocity, since typically it holds $|\mathbf{q}| \gg |\mathbf{Q}|$ and the dispersion of graphene is linear in the investigated $\mathbf{q}$-region for coherent or even thermalized TMD excitons. 
The remaining terms in eqs. (\ref{GrapheneCoherence}) and (\ref{ExcitonicCoherence}) result from the F\"orster coupling between both monolayers. The coupled eqs. (\ref{GrapheneCoherence}) and (\ref{ExcitonicCoherence}) describe new quasiparticles from electron-hole pairs in graphene and excitons in TMDs. The most direct approach in the weak coupling limit is to perform a Born-Markov approximation\cite{Kuhn1992,Schilp1994,Scully} for the polarization in graphene $R^\mathbf{q}$. This approximation is valid as long as the F\"orster induced interlayer coupling is small compared to the spectral bandwidth provided by the broad electronic energy distribution in graphene. The Born Markov approximation carried out is formally equivalent to a Wigner-Weisskopf approximation where the graphene continuum acts as a bath\cite{Scully}. This way, we have access to the dephasing rate and the energy renormalization for excitons in the TMD due to F\"orster coupling with the graphene layer:
\begin{align}
\partial_t P^{\nu}_\mathbf{Q}=-\frac{i}{\hbar}(E^{\nu}_\mathbf{Q}+\Delta_\mathbf{Q}^{\nu}-i\Gamma_\mathbf{Q}^{\nu}) P^{\nu}_\mathbf{Q}.
\end{align}

For the broadening we obtain

\begin{align}
\Gamma_\mathbf{Q}^{\nu}= |V(\mathbf{Q},\mathbf{z},\nu)|^2 \pi  \sum_\mathbf{q} \delta (E^{\mathbf{q}}-E^{\nu}_\mathbf{Q}) \label{dephasing_0}  \\
=\frac{ d_T^2 d_G^2 | \varphi^{\nu}( \mathbf{r}=0 ) |^2 E_\mathbf{Q}^{\nu} }{64 \pi^2  \epsilon^2_0 \epsilon^2  \hbar^2 v_F^2} 
 |a(\mathbf{Q},\mathbf{z})|^2, \label{dephasing}
\end{align}
while the energy renormalization reads

\begin{align}
\Delta^{\nu}_\mathbf{Q}&=- |V(\mathbf{Q},\mathbf{z},\nu)|^2 \sum_\mathbf{q} \frac{1}{E^\mathbf{q}-E^{\nu}_\mathbf{Q}}  \label{lineshift_0}   \\
&=-\frac{d_T^2 d_G^2 |\varphi^{\nu} (\mathbf{r}=0)|^2 q_{max}}{32 \pi^3 \epsilon_0^2 \epsilon^2  \hbar v_F  } \times \nonumber \\ &\times\left(1+\frac{E_\mathbf{Q}^{\nu}}{2 \hbar v_F q_{max}} ln|1-\frac{2 \hbar v_F q_{max}}{E_\mathbf{Q}^{\nu}}|\right) |a(\mathbf{Q},\mathbf{z})|^2  \label{lineshift}
\end{align}

Here, we neglect the F\"orster interaction mediated influence of different TMD excitons on each other. For an optical excitation, we have $E^{\nu}_\mathbf{Q} \approx \hbar \omega_{opt}$, where $\omega_{opt}$ is the incident laser frequency. Note that we added a factor of $2$ to take account to the spin degree of freedom in graphene. 
Note that the momentum sum appearing in the expression for the lineshift, equation \ref{lineshift_0}, in general diverges. This problem is also known from the computation of the Lambshift, where the atomic states couple to the mode continuum of the radiational field which results in a divergent self energy contribution\cite{Scully}. Similar, in the current case, the energy renormalization does depend on the choice of a physically motivated momentum cutoff $q_{max}$. Therefore we do not compute the energy renormalization explicitly but will discuss it qualitatively in the following.
We find that the dephasing rate, eq. (\ref{dephasing}), and the energy renormalization, eq. (\ref{lineshift}), show the same momentum and layer separation dependence, since both are proportional to $a(\mathbf{Q},\mathbf{z})|^2$. 

The appearing function $|a(\mathbf{Q},\mathbf{z})|^2$ in eqs. \ref{dephasing} and \ref{lineshift} has to be evaluated.  $a(\mathbf{Q},\mathbf{z})$ can be integrated by Schwinger parameterizing the denominator of the integrand \cite{Prausa2017}
\begin{equation}
\frac{1}{x^{k}} = \frac{1}{\Gamma (k)} \int_{0}^{\infty} dt t^{k-1} e^{-tx},
\end{equation}
with $x,k\in \mathbb{R}$ and $\Gamma (k)$ denoting the Gamma function. Inserting this expression in the integral $a( \mathbf{Q},z)$ the $d^2 s$ integration can be performed straight forward. Also the $dt$ integration over hermite gaussian polynomials
\begin{equation}
a(\mathbf{Q},z)=2 \pi \frac{\mathbf{e}_1 \cdot \mathbf{Q}\,\, \mathbf{e}_2 \cdot \mathbf{Q}}{Q} e^{-Qz}\label{aaa}
\end{equation}
Our result coincides nicely with the result given in reference \onlinecite{Batsch1993}. Note that we obtain a different expression compared to reference \onlinecite{Tomita1996}.
Since we are interested in the general behavior of the F\"orster rate, the dependence on the angle of the center of mass momentum is averaged out, which yields
\begin{equation}
a(Q,z) = \pi Q e^{-Qz} \mathbf{e}_1 \cdot \mathbf{e}_2 
\end{equation}

The last step is to square this expression and sum it over $K$ and $K'$ point in graphene. The latter is implicitly included in the summation over the momentum $\mathbf{q}$ in equation \ref{lineshift_0} and \ref{dephasing_0}. Note that we have to sum over two orthogonal dipole moments in graphene \cite{Stroucken2011}, and therefore the angular dependence in the final result drops. We obtain the final expression 
\begin{equation}
|a(Q,z)|^2= \pi^2 Q^2 e^{-2Qz}.
\end{equation}
Using eqs. \ref{dephasing} and \ref{lineshift}, the corresponding dephasing rates and energy renormalizations can be evaluated quantitatively.

\section{Results}

We exploit the derived equations to calculate F\"orster induced dephasing rate for the case of a van der Waals heterostructure consisting of graphene and monolayer tungsten disulfide (WS$_2$) as an exemplary TMD on a quartz substrate ($\epsilon=$3.9). In particular, we exploit Eqs. (\ref{dephasing})  with parameters given in Table \ref{parameters} and compute the dephasing rate in the WS$_2$ monolayer in the presence of the graphene layer. 

\begin{table}[b!] 
\centering 
\caption{Parameters used in the computation. $^*$ determined numerically by using the method given in \cite{Berghauser2014,Selig2016}} 
\begin{tabular}{l l | l l l}\hline 
	Param. &  & Param. & & Ref. \\ 
   $\hbar$ & \unit[0.658]{eV fs}  & $d_G$ & \unit[0.25]{e nm}  & \cite{Winzer2011} \\ 
   $e$ & \unit[1]{e} &  $v_F$ & \unit[1]{nm fs$^{-1}$} & \cite{CastroNeto2009}\\
   $\epsilon_0$ & \unit[5.5$\cdot$10$^{-2}$]{e$^2$eV$^{-1}$nm$^{-1}$}  & $d_{WS_2}$ & \unit[0.4]{e nm}  &\cite{Selig2016}\\ 
   $k_B$ & \unit[8.6$\cdot$10$^{-5}$]{eV K$^{-1}$}   & $|\varphi_{WS_2}(\mathbf{r}=0)|$ & \unit[0.49]{nm$^{-1}$} & $^*$\\
   $\epsilon_{SiO_2}$ & \unit[3.9]{}  & $E^{1s}_{WS_2}$ & \unit[2.0]{eV} & \cite{Christiansen2017} \\
    & & $M_{WS_2}$ & \unit[3.5]{eVfs$^2$nm$^{-2}$} & \cite{Kormanyos2015} \\ 

   \hline

\end{tabular} \label{parameters}
\end{table}

\begin{figure}[t!]
 \begin{center}
\includegraphics[width=0.9\linewidth]{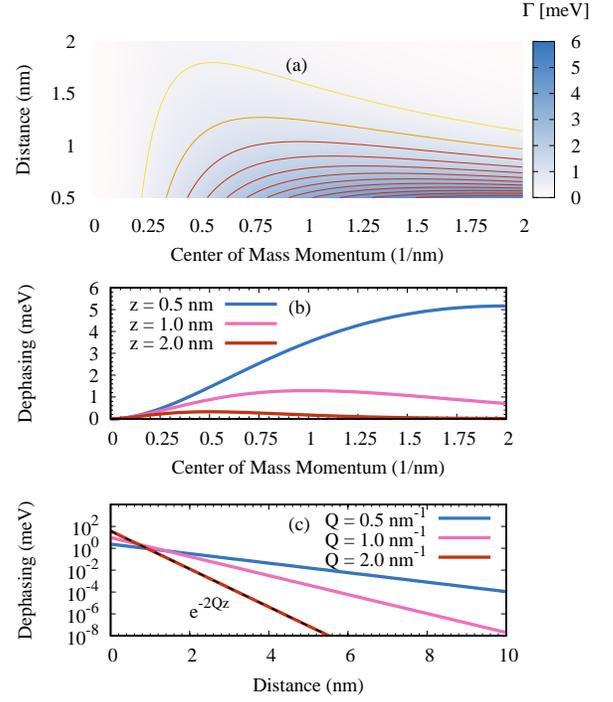}
 \end{center}
 \caption{\textbf{F\"orster induced broadening and energy renormalization of excitonic resonances.} (a) Surface plot of the dephasing rate, eq. \ref{dephasing} as a function of the excitonic center-of-mass momentum and the distance between the graphen and the TMD layer.  Dephasing rate, eq. \ref{dephasing}, (b) as a function of the center-of-mass momentum of the TMD exciton for fixed layer distances and (c) as a function of the interlayer distance for fixed exciton momenta.}
 \label{fig_dephasing}
\end{figure}

In Figure \ref{fig_dephasing} (a), we show a contour plot of the F\"orster induced dephasing rate as a function of the center-of-mass momentum $\mathbf{Q}$ of the TMD exciton and the distance between the TMD and the graphene layer. We find dephasing rates ranging from more than \unit[5]{meV} at $\mathbf{Q}=$\unit[2]{nm$^{-1}$} and $z=$\unit[0.5]{nm} (relevant for the generation of incoherent excitons for photoluminescence \cite{Thranhardt2000}), to \unit[0]{meV} for $\mathbf{Q}=0$ (relevant for coherent optical absorption) and $z=$\unit[0.5]{nm} (corresponding to the situation where the WS$_2$ layer and the graphene layer are closely stacked\cite{Hill2017}). As expected, the dephasing rate decreases as a function of the interlayer spacing. 
 
To investigate the dependence on the center-of-mass momentum in more detail, we show cuts at different interlayer separations of \unit[0.5]{nm},  \unit[1.0]{nm} and  \unit[2]{nm} in Fig. \ref{fig_dephasing} (b). We find for all interlayer separations an initial increase of the F\"orster rate followed by an exponential decay, which can be directly extracted from eq. (\ref{aaa}). 
Excitons, visible in coherent optical experiments (without any scattering or thermalization), exhibit vanishing center-of-mass momenta $\mathbf{Q} \approx 0$, due to momentum conservation \cite{Kira2006,Kochbuch,Berghauser2014}. Our computations predict no impact of the F\"orster coupling on the linewidth and resonance position of coherent excitons at $\mathbf{Q}=0$. In reference \onlinecite{Hill2017}, H. Hill and co-workers found an additional broadening of the WS2$_2$ resonance in a coherent reflectance measurement of about \unit[5]{meV} compared to the monolayer. Since the F\"orster dephasing rate vanishes for $\mathbf{Q}=0$ it can be ruled out as a direct source of the peak broadening which requires ongoing investigations.

In our computations, we find a strong impact of the dielectric constant of the enviroment due to the screening of the Coulomb potential and exciton wavefunction. The latter effect turns out to be less significant. To illustrate this, we discuss briefly two limits. Exemplary, for a free standing heterostructure ($\epsilon=$1) with an interlayer separation of \unit[0.5]{nm} we find a F\"orster induced dephasing of about \unit[10]{meV} at $Q=$\unit[2]{nm$^{-1}$}. In contrast, for a heterostructure encapsulated by hexagonal boron nitride ($\epsilon=$7) we find a F\"orster rate of approximately \unit[1.5]{meV} at $Q=$\unit[2]{nm$^{-1}$}. 

\begin{figure}[t!]
 \begin{center}
\includegraphics[width=0.9\linewidth]{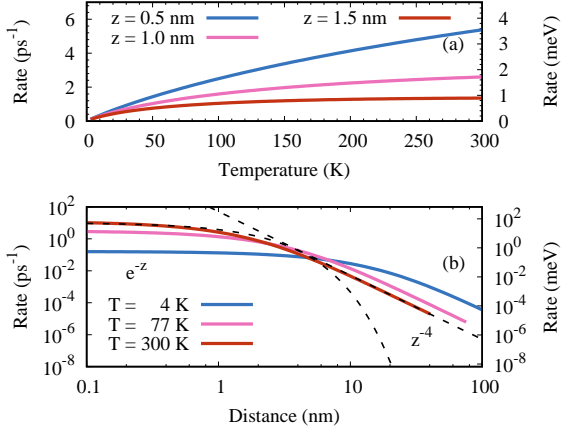}
 \end{center}
 \caption{\textbf{Transition Rate} (a) Transition rate, eq. \ref{dephasing} and \ref{thermal}, as a function of the temperature of the exciton in the TMD for 3 different layer distances. (b) Transition rate as a function of the interlayer distance for 3 different exciton momenta.}
 \label{fig_rate}
\end{figure}

Next we investigate the dependence of the F\"orster induced dephasing rate on the interlayer separation. In Fig. \ref{fig_dephasing} (c) we show the dephasing rate as a function of the interlayer distance for three different center-of-mass momenta $Q$. First, we find for all  $Q$ a decreasing behavior. For elevated center of mass momenta, the F\"orster induced dephasing rate exhibits an exponential decay which is depicted in figure \ref{fig_dephasing} (c).

To get a first impression of the F\"orster transition rate of a thermalized exciton distribution in the WS$_2$ monolayer into the graphene layer, we apply a thermal average. In the low excitation limit the equation for the incoherent exciton density  $N_\mathbf{Q} = \delta \langle P^{\dagger}_\mathbf{Q} P_\mathbf{Q} \rangle$ \cite{Selig2018}  reads : $\partial_t N_\mathbf{Q} = - 2 \Gamma_\mathbf{Q} N_\mathbf{Q}$. Here, supported by the fast relaxation within the graphene layer \cite{Winzer2010}, the occupation in graphene was assumed to vanish. Under the assumption of an initially thermalized exciton Boltzmann distribution $N_\mathbf{Q} = N e^{-\frac{E_\mathbf{Q}}{k_B T}}$ in the TMD, the temperature dependent transition rate can be obtained by thermally averaging the momentum dependent transition rate  $\Gamma_T=2 \langle \Gamma_\mathbf{Q}^{1s} \rangle_T$.
 This assumption is justified by the fast exciton-phonon scattering rates which mediate the thermalization, being in the order of some \unit[10]{fs}\cite{Selig2018}.  
The corresponding thermal average of the function $|a(T,z)|^2=\langle |a(\mathbf{Q},z)|^2 \rangle_T$ reads

\begin{align}
|a(T,z)|^2 =\frac{\pi^2}{2} \frac{1}{\lambda^6} \bigg(2 \lambda(T)^2 (\lambda(T)^2+z^2)\nonumber \\-e^{z^2/\lambda(T)^2} \lambda \sqrt{\pi} z (3\lambda(T)^2+2z^2)erfc(z/\lambda(T))\bigg) \label{thermal}
\end{align}

with the error function $\rm{erfc}(z/\lambda(T))$ and the thermal wavelength $\lambda(T)=\frac{\hbar}{\sqrt{2 M k_B T}}$. Interestingly, in the second term a high numerical accuracy is required in the evaluation since the function $e^{a^2}$ is fast increasing and the function $erfc(a)$ is fast decreasing as a function of $a$ \cite{Tomita1996}.

Figure \ref{fig_rate} shows the transition rate as a function of temperature for three selected interlayer separations $z$. We find for all $z$ an increase of the rate with increasing temperature. Exemplary, we predict a rate of \unit[5]{ps$^{-1}$} for a closely stacked heterostructure ($z=$\unit[0.5]{nm}) at room temperature. A similar time scale was reported for the transition of carriers from WS$_2$ to graphene for a closely stacked heterostructure from non-linear differential reflectance contrast measurements (pump-probe)\cite{HeJiaqi2014}. The incoherent exciton occupation is visible in coherent reflectance contrast measurements as a bleaching of the excitonic polarization $i \hbar \partial_t P_\mathbf{0} = E P_\mathbf{0} + (1 - 2 \sum_\mathbf{Q} \Xi_\mathbf{Q} N_\mathbf{Q}) \mathbf{d} \cdot  \mathbf{E}_T$, with the excitonic form factor $\Xi_\mathbf{Q} = \sum_\mathbf{q} | \varphi_\mathbf{q} |^2 | \varphi_\mathbf{q+\frac{1}{2} Q} |^2$ during and after the thermalization \cite{Hawton1998,Katsch2018}. 
Our computation reveals a value of \unit[4]{ps$^{-1}$} for a heterostructure consisting of monolayer MoSe$_2$ and graphene (obtained with parameters from table \ref{parameters_mose2}). This value also coincides qualitativley with the recently reported exciton lifetime of about \unit[1]{ps} in a MoSe$_2$-graphene stack\cite{Froehlicher2018}.

\begin{table}[b!] 
\centering 
\caption{Parameters for MoSe$_2$ as TMD material used in the computation. $^*$ determined numerically by using the method given in \cite{Berghauser2014,Selig2016}} 
\begin{tabular}{ l l l}\hline 
	 Param. & & Ref. \\ 
   $d_{MoSe_2}$ & \unit[0.25]{e nm}  &\cite{Selig2016}\\ 
   $|\varphi_{MoSe_2}(\mathbf{r}=0)|$ & \unit[0.64]{nm$^{-1}$} & $^*$\\
   $E^{1s}_{MoSe_2}$ & \unit[1.6]{eV} & \cite{Christiansen2017} \\
   $M_{MoSe_2}$ & \unit[6.1]{eVfs$^2$nm$^{-2}$} & \cite{Kormanyos2015} \\ 

   \hline

\end{tabular} \label{parameters_mose2}
\end{table}

In figure \ref{fig_rate} (a), the increasing F\"orster induced transition rate as a function of temperature can be explained as follows: At low temperatures, the exciton thermalizes in narrow Boltzmann distributions at very small momenta with a width of $kT$. For this distribution we find a vanishing F\"orster rates, because the F\"orster coupling as a function of the center of mass momentum $\mathbf{Q}$ vanishes for $\mathbf{Q}=0$. 
Increasing the temperature leads to a broadening of the exciton distribution in momentum space. The leads to the occupation of exciton states with larger F\"orster rates, cf. Fig. \ref{fig_dephasing} (a) and (b). This results in a increase of the F\"orster transition rate at elevated temperatures. However, since even at room temperature most of the excitons are located at low momenta, the transition rate is still increasing at elevated temperatures. 

Finally, in Fig.  \ref{fig_rate} (b), we show the transition rate as a function of the interlayer separation for three different temperatures. For small distances we find a $e^{-z}$ law for all investigated temperatures, which is consistent with the observations for the dephasing rate. At larger interlayer spacings we find a $z^{-4}$ law, which is consistent with the limits of the function in equation \ref{thermal}. The observed $z$ dependence is consistent with the observations for F\"orster rates in porpyrin-functionalized graphene. \cite{Malic2014} The developed theoretical approach can be applied to other van der Waals heterostructures including multilayers of the same TMD or stacks of different TMD monolayers\cite{Ovesen2018}.

\section{Conclusion}

In conclusion, we presented a simple analytic model describing the F\"orster mechanism in van der Waals heterostructures consisting of graphene and a monolayer TMD. We predict F\"orster rates leading to a fast energy transfer between the TMD and graphene on a picosecond timescale. Exemplary the F\"orster induced transition rate of thermalized WS$_2$ excitons is about \unit[5]{ps$^{-1}$} at room temperature. This was found to nicely coincide with recent pump probe \cite{HeJiaqi2014} and photoluminescence \cite{Froehlicher2018} measurements. So far, we have not included recently investigated dark exciton states with momenta far beyond the light cone\cite{Qiu2015,Wu2015,Selig2016,Selig2018}, which will be addressed in future work.

\section*{Acknowledgement}
In particular we thank the referee of the manuscript for a fruitful discussion. The authors gratefully acknowledge stimulating discussions with Gunnar Bergh\"auser (Chalmers, G\"oteborg), Manuel Kraft and Florian Katsch (TU Berlin). This work was funded by the Deutsche Forschungsgemeinschaft (DFG, German Research Foundation), Projektnummer 182087777–SFB 951 (projects A8 and B12, N.K. and A.K.) and Projektnummer 43659573–SFB 787 (project B1, M.S.). This project has also received funding from the European Unions Horizon 2020 research and innovation programme under Grant Agreements No. 696656 (Graphene Flagship, E.M.) and No. 734690 (SONAR, A.K.). Further we acknowledge funding
from the Swedish Research Council and Stiftelsen Olle Engkvist (E.M.). This work was partially supported by the National Research Foundation of Korea (NRF) Grants funded
by the Korean Government (MSIP) (2018R1A2B6001449).

\bibliographystyle{apsrev4-1}

\end{document}